\documentstyle[12pt]{article}
\begin{document}
\baselineskip=24pt
\newcommand{\Lam}{\Lambda_{\scriptscriptstyle {\rm \overline{MS}} }}
\newcommand{\Lamm}[1]{\Lambda
{\vspace{-2pt} \scriptscriptstyle ^{^{\rm #1}}
    _{\rm \overline{MS}} }}
\newcommand{\dsp}{\displaystyle}
\newcommand{\dfr}[2]{ \displaystyle\frac{#1}{#2} }
\newcommand{\Lag}{\Lambda \scriptscriptstyle _{ \rm GR} }
\newcommand{\pa}{p\parallel}
\newcommand{\pe}{p\perp}
\newcommand{\pet}{p\top}
\newcommand{\paa}{p'\parallel}
\newcommand{\pee}{p'\perp}
\newcommand{\pete}{p'\top}
\renewcommand{\baselinestretch}{1.5}
\begin{titlepage}
\vspace{-20ex}
\begin{flushright}
\vspace{-3.0ex}
    \it{AS-ITP-94-11} \\
\vspace{-2.0mm}
       \it{March, 1994}\\
%\vspace{-2.0mm}
\vspace{5.0ex}
\end{flushright}
\centerline{\Large %\sf
The forward-backward asymmetry with $Z'$ effects}
\centerline{\Large %\sf
in the process $e^{+}$ $e^{-}$ $\rightarrow$ $\mu^{+}$ $\mu^{-}$}
\vspace{6.4ex}
\centerline{\large %\sf
Ma Wen-Gan, Sun La-Zhen, Liu Yao-Yang and Jiang Yi}
\centerline{\small Modern Physics Deptment, University of Science and
Technology of China, Anhui 230027, China}
\vspace{3ex}
\centerline{\large %\sf
Chang Chao-Hsi}
\centerline{\small CCAST (World Laboratory), P.O.Box 8730,
Beijing 100080, China}
\centerline{\small Institute of Theoretical Physics, Academia Sinica,
P.O.Box 2735, Beijing 100080, China\footnote{Mailing address.}}
\vspace{3ex}
\begin{center}
\begin{minipage}{100mm}
\centerline{\large %\sf
Abstract}
\vspace{1.5ex}
\small
{Having taken the QED radiative correction and one-loop weak corrections
into account, we have computed the unpolarized forward-backward
asymmetry in the process $e^+e^- -> \mu^+\mu^-$
at a very reasonable accuracy. Special attention to
the effects around the LEP 200 energy,
induced by the possible $Z'$ boson of the left-right symmetric model
$SU(2)_{L} \times SU(2)_{R} \times U(1)_{B-L}$, is paid. The
numerical results of the directly measurable
asymmetry $A_{FB}^{\mu}$ versus the CMS energies are presented in figures.
One may be convinced by the results that precise measurements
on the asymmetry in future colliders will possibly present some indirectly
evidence of the $Z'$ boson.}
\vskip 3mm
{PACS number(s): 13.15.Jr, 13.40.Ks, 14.80.Er, 14.60.Cd}
\end{minipage}
\end{center}
\end{titlepage}

\newpage

Up to now there has been no evidence which shows any discrepancy between
the predictions of the standard model $SU_{c}(3) \times SU_{L}(2)
\times U_{Y}(1)$
(SM) and the measurements at the nowadays best experimental accuracy.
Nevertheless, the SM still should be considered as a low energy limit of a
more fundamental theory, such as a unification of the electroweak and strong
interactions at much higher energy scale,
due to the fact that there are so many
parameters in the model, which should be understood. Being one of the efforts
in the direction, the left-right symmetric model $SU(2)_{L}
\times SU(2)_{R} \times U(1)_{B-L}$
(LRM) is a comparative natural extension of the SM and has some
predictabilities.
An additional neutral gauge boson $Z'$ is one of its important predictions.
If favorable parameters are taken by nature,
the boson may have not too heavy mass, so that it is
accessible at future colliders[1]. The neutral gauge boson $Z'$ is to couple
to the fermion-antifermion pair so must effect the measurable
forward-backward
asymmetry in fermion pair to fermion pair processes. In this report,
we will focus lights on the effects due to the boson.
Pursuing the accuracy of a few percents,
which may be accessible in theoretical calculations and experimental
measurements both. The theoretical calculations to the accuracy
just correspond up to the QED radiative correction and
the electroweak one-loop correction level.
We are to analyze the unpolarized forward-backward asymmetry
for muons $(A^{\mu}_{FB})$ in the $e^{+}$ $e^{-}$ $\rightarrow$
$\mu^{+}$ $\mu^{-}$ process at the accurate level in two extreme
cases of LRM in the paper. To see the effects clearly we also compare
the results with those obtained from
SM theory. Note that in the calculations,
the most recent constraints on the $Z'$ mass by experiments[2] have been
concerned.
The results of our analysis are presented
by figures,the measurable asymmetry
$(A^{\mu}_{FB})$ versus the CMS energies of collisions. As expected, the QED
radiative correction is important and smears the asymmetry effects strongly.
However our results still show certain possibilities that one may find some
evidence of the new gauge boson $Z'$ at LEP 200 or at future $e^{+} e^{-}$
colliders if analyzing the measured $A^{\mu}_{AB}$ against the calculated
one carefully. How significant the evidence to be depends on the value
of the $Z'$ mass (if it exists) and the
accuracy of the experimental measurements
on the asymmetry as well.

In the LRM a right-handed $SU(2)_{R}$ gauge interaction is extended from
SM. Although in principle a mixing between $W^{\pm}_{L}$ and $W^{\pm}_{R}$ in
LRM is possible, we suppose it can be neglected, that is
indicated by experiments e.g.
the charge currents of quarks and leptons
are of $V-A$, as indicated by experiments
within the accuracy of measurements on them. In the model the possible
right-handed neutrinos are required very heavy
by experimental measurements too,
and the mixing between them and the
left-handed ones is so tiny that it can do
nothing what we are interested in here.
The parameter $\alpha_{LR}$ is used to
describe the couplings of the heavy boson to fermions.
\begin{eqnarray*}
   \alpha_{LR} = \sqrt{ \frac{ \cos^2{\theta_{w}} }
   { \lambda^2\cdot \sin^2{\theta_{w}} }-1}.
\hskip 6cm (1)
\end{eqnarray*}
We define $\lambda=g_{L}/g_{R}$, $g_{L}$ and
$g_{R}$ represent the $SU(2)_{L,R}$
coupling constants. When $\alpha_{LR}$ may take the value
either $\sqrt{\frac{2}{3}}$ or 1.53, which corresponds to
its lowest and upper bounds respectively. For $\lambda=1$
(i.e. $\alpha_{LR}\simeq 1.53$ for $\sin^{2}{\theta_{w}}=0.23$),
it is just the case often used in LRM, while when $\alpha_{LR}=\sqrt{2/3}$,
the LRM is identical to the $\chi$ model,
attributed from the superstring inspired $E_{6}$ gauge model[3].
In the following we will compute the effects in
the two extreme cases.

  The neutral-current lagrangian in LRM is given by[4]
\begin{eqnarray*}
   L = igJ_{3L}W_{3L} + i\frac{g}{\lambda}J_{3R}W_{3R} + ig'J_{B-L}B
\hskip 4.5cm (2)
\end{eqnarray*}
where
\begin{eqnarray*}
   g=g_{L}=\frac{e}{\sin{\theta_{w}}}. \hskip 9cm (3)
\end{eqnarray*}
Here we would define
\begin{eqnarray*}
 y = \sqrt{\frac{\cos^2{\theta_{w}}}{\lambda^2}-\sin^2{\theta_{w}}}
\hskip 7.5cm (4)
\end{eqnarray*}
for later usages.
$W_{3L},W_{3R}$ and $B$ are the
$SU(2)_{L},SU(2)_{R}$ and $U(1)_{B-L}$ gauge
fields; $g,g/\lambda$ and $g'$ are the coupling
constants of gauge fields to
the fermion currents:
\begin{eqnarray*}
   J^{\mu}_{3L} = \bar{\psi}\gamma^{\mu}I_{3L}\psi,
   \hskip 0.5cm J^{\mu}_{3R} = \bar{\psi}\gamma^{\mu}I_{3R}\psi
\hskip 5cm (5)
\end{eqnarray*}
\begin{eqnarray*}
   J^{\mu}_{B-L} = \bar{\psi}\gamma^{\mu}{\frac{B-L}{2}}\psi \hskip 6.5cm (6)
\end{eqnarray*}
As $Z_{L}$, $Z_{R}$ are not the mass eigenstates in general,
the corresponding mass eigenstates $Z^{0}$
and $Z'$ are of a mixture of them as follows:
\[
\left( \begin{array}{ll} Z&\\ Z'& \end{array}\right)=
\left( \begin{array}{ll} \cos\phi & \sin\phi \\-\sin\phi & \cos\phi
      \end{array}\right)
\left( \begin{array}{ll} Z_{L}& \\ Z_{R}& \end{array}\right) \hskip 3.6cm (7)
\]
{}From equation (2), (5), (6) and (7), we may read out
the vertices of $\gamma\mu\mu$, $Z^{0}\mu\mu$ and $Z'\mu\mu$ etc
are in the form $ig\gamma^{\mu}(g^{(V)}_{\cdots}-g^{(A)}_{\cdots}
\gamma_{5})$[6], where
\begin{eqnarray*}
   g^{(V)}_{\gamma\mu\mu} = -\sin\theta_{w},
\hskip 9cm (8a)
\end{eqnarray*}
\begin{eqnarray*}
   g^{(A)}_{\gamma\mu\mu}= 0,
\hskip 10cm (8b)
\end{eqnarray*}
\begin{eqnarray*}
   g^{(V)}_{Z\mu\mu} &=& {\frac{1}{4}}[(-\cos\theta_{w}+3\sin\theta_{w}
\tan\theta_{w})\cos\phi+(-{\frac{y}{\cos\theta_{w}}}+{\frac{2
\sin\theta_{w}\tan
\theta_{w}}{y}})\sin\phi] \\
         &=&  \frac{1}{4 \cos\theta_{w}}[V_{1} \cos\phi+V_{2} \sin\phi],
\hskip 5.5cm (9a)
\end{eqnarray*}
\begin{eqnarray*}
   g^{(A)}_{Z\mu\mu} = \frac{1}{4\cos\theta_{w}}(-\cos\phi+y \sin\phi),
\hskip 5.5cm (9b)
\end{eqnarray*}
\begin{eqnarray*}
   g^{(V)}_{Z'\mu\mu} &=&
    \frac{1}{4}[(-\frac{y}{\cos\theta_{w}}+
\frac{2 \sin\theta_{w} \tan\theta_{w}}{y})\cos\phi-(-\cos\theta_{w}+
3\sin\theta_{w} \tan\theta_{w}) \sin\phi] \\
&=& \frac{1}{4 \cos \theta_{w}} (V_{2} \cos \phi- V_{1} \sin \phi),
\hskip 5.3cm (10a)
\end{eqnarray*}
\begin{eqnarray*}
   g^{(A)}_{Z'\mu\mu} =\frac{1}{4\cos\theta_{w}}[y \cos\phi+\sin\phi],
\hskip 6cm (10b)
\end{eqnarray*}
and with
$$ V_{1}=4 \sin^{2}\theta_{w}-1, \;\;\;
V_{2}=\displaystyle\frac{2\sin^{2}\theta_{w}}{y}-y .$$

The typical Feynman diagrams which contribute to
the process $e^{+} e^{-} \rightarrow \mu^{+}
\mu^{-}$ at the tree level in LRM are
shown in Fig. 1 . The differential cross
sections in Born approximation in the framework of SM is
given in references[4,5]. Here to calculate out the differential cross
section in LRM is necessary. At tree level and neglecting the external
fermion masses, we may write the cross section in the same
way as in SM, i.e. both are in
the form as Eq.(11) of Refs.[6,7]:
\begin{eqnarray*}
   \frac{d\sigma^{SM,LRM}}{d\Omega} =
\frac{\alpha^{2}}{4s}{[(1-z^{2}) G_{1}^
{SM,LRM}(s)+2 z G_{3}^{SM,LRM}(s)}],  \hskip 3cm (11)
\end{eqnarray*}
here $z=\cos\theta$, only the functions, $G_{1}(s)$ and $G_{3}(s)$,
are different from each other for SM and LRM. They have
the formulae as follows respectively:
\begin{eqnarray*}
   G_{1}^{SM}(s) = 1+2 v_{e} v_{\mu} Re(D_{Z}(s))+(v^{2}_{e}+a^{2}_{e})
(v^{2}_{\mu}+a^{2}_{\mu}) |D_{Z}(s)|^{2},
\hskip 1.5cm (12a)
\end{eqnarray*}
\begin{eqnarray*}
   G_{3}^{SM}(s) = 2 a_{e} v_{\mu} Re(D_{Z}(s))+4 v_{e} a_{e} v_{\mu} a_
{\mu} |D_{Z}(s)|^{2},  \hskip 3.5cm (12b)
\end{eqnarray*}
for those of SM, and
\begin{eqnarray*}
   G_{1}^{LRM}(s) &=&
1+\frac{B}{8}\cdot\frac{s-M_{Z}^{2}}{s}\cdot|D_{Z}(s)|^2
   (V_{1}\cos\phi+V_{2}\sin\phi)^{2} \\
&& +\frac{B^{2}}{256} |D_{Z}(s)|^{2}[(V_{1}\cos\phi+V_{2}\sin\phi)^{2}+
(-\cos\phi+y\sin\phi)^{2}]^{2}  \\
&& +\frac{B^{2}}{256} |D_{Z'}(s)|^{2}[(V_{2}\cos\phi
-V_{1}\sin\phi)^{2}+(y\cos\phi+\sin\phi)^{2}]^{2}  \\
&& +\frac{B(s-M_{Z'}^{2})}{8s}
|D_{Z'}(s)|^{2}(V_{2}\cos{\phi}-V_{1}\sin{\phi})^{2}  \\
&& +\frac{B^{2}}{128}\frac{(s-M_{Z}^{2})}{s}\cdot\frac{(s-M_{Z'}^{2})}{s}
\cdot|D_{Z}(s)|^2\cdot|D_{Z'}(s)|^{2}\\
&&\cdot[(V_{1}\cos{\phi}
+V_{2}\sin{\phi}) \cdot(V_{2}\cos{\phi}-V_{1}\sin{\phi})\\
&&+(-\cos{\phi}+y\sin{\phi})(y\cos{\phi}+\sin{\phi})]^{2}, \hskip 3.8cm (13a)
\end{eqnarray*}
\begin{eqnarray*}
   G_{3}^{LRM}(s) &=& \frac{B}{8}\cdot\frac{s-M_{Z}^{2}}{s}\cdot|D_{Z}(s)|^2
    (-\cos\phi+y\sin\phi)^{2} \\
&& +\frac{B^{2}}{64} |D_{Z}(s)|^{2}(V_{1}\cos\phi+V_{2}\sin\phi)^{2}
(-\cos\phi+y\sin\phi)^{2} \\
&& +\frac{B^{2}}{64}
|D_{Z'}(s)|^{2}(V_{2}\cos\phi-V_{1}\sin\phi)^{2}(y\cos\phi
+\sin\phi)^{2}   \\
&& +\frac{B(s-M_{Z'}^{2})}{8s} |D_{Z'}(s)|^{2}
(y\cos\phi+\sin\phi)^{2} \\
&& +\frac{B^{2}}{128} \frac{s-M_{Z}^{2}}{s}\cdot\frac{s-M_{Z'}^{2}}{s}
\cdot|D_{Z}(s)|^2
\cdot|D_{Z'}(s)|^{2}\\
&&\cdot[(V_{1}\cos\phi+V_{2}\sin\phi)(y\cos\phi+\sin\phi) \\
&& +(-\cos\phi+y\sin\phi)
(V_{2}\cos\phi-V_{1}\sin\phi)]^{2}, \hskip 3.8cm (13b)
\end{eqnarray*}
for those of LRM, here we have taken the notation:
\begin{eqnarray*}
   B = \frac{1}{\sin^{2}\theta_{w}\cos^{2}\theta_{w}}, \hskip 11cm (14a)
\end{eqnarray*}
\begin{eqnarray*}
   D_{Z}(s) = \frac{s}{s-M_{Z}^{2}+iM_{Z}\Gamma^{0}_{Z}}, \hskip 8.8cm (14b)
\end{eqnarray*}
\begin{eqnarray*}
   D_{Z'}(s) =
\frac{s}{s-M_{Z'}^{2}+iM_{Z'}\Gamma^{0}_{Z'}}. \hskip 9.2cm (14c)
\end{eqnarray*}

According to the definition, the unpolarized forward-backward
asymmetry $A_{FB}^{\mu}$ is properly related to the cross section:
\begin{eqnarray*}
   A_{FB}^{\mu} &=&
\frac{2\pi(\int_{0}^{\frac{\pi}{2}}\frac{d\sigma}{d\theta}
d\theta - \int_{\frac{\pi}{2}}^{\pi}\frac{d\sigma}{d\theta}d\theta)}{\sigma
_{T}} \\
&=& \frac{\sigma_{FB}}{\sigma_{T}}. \hskip 6.5cm (15)
\end{eqnarray*}
In Born approximation, the $A_{FB}^{\mu}$ is formulated simply
\begin{eqnarray*}
   A_{FB}^{\mu} = \frac{3}{4} \cdot
\frac{G_{3}^{SM,LRM}(s)}{G_{1}^{SM,LRM}(s)}.
\hskip 6cm (16)
\end{eqnarray*}

It is known that the mass of neutral
boson $Z'$ in LRM is located far beyond the
energy range of the colliders LEP-1 and LEP 200 [2],
therefore the effects from
the $Z'$-exchange diagrams cannot be expected
to be great. In order to detect the expected small $Z'$
effects from quite great background of SM, we need to calculate
the background to a higher order level precisely.
The weak corrections up to one-loop
in SM is needed here. The corrections in SM, including
propagator and vertex corrections as
well as the box diagram contributions, are given by Wolfgang F. L. Hollik
in Ref.[5].
To the accurate level, the differential cross section
still may write in the form as Eq.(11), but the functions
$G_{1}$ and $G_{3}$ now will depend on the two Mandelstam variables
$t$ as well as $s$, thus when we consider the one-loop weak corrections of
SM alone to the cross section, we may just simply use the
invariant functions $G_{1}(s,t)$ and
$G_{3}(s,t)$ specified in the equation (6.34) of Ref.[5] and to
replace them properly into
the above formula Eq.(11). As for the radiative corrections of LRM,
since the $Z'$ mass is probably much larger than the energy range
of LEP 200, where we are considering in the paper,
the $Z'$-exchange contribution is rather small, so that it is
accurate enough to neglect safely the $Z-Z'$ mixing, the
interference between the pure weak loop correction and the $Z'$-exchange
diagram and the contribution from those loops contained one or more
$Z'$ boson propagators as well at this moment. The
extra Higgs sector correction in LRM is also neglected for simplification.

However, the QED radiative corrections, especially those from
the colinear and/or soft
photon emissions, are great due to a tiny mass of electrons.
Additionally it makes that the precise measured
value of the asymmetry depends on
the experimental conditions. As
expected, among all the QED corrections, those
of the initial states are the largest[8]. We use the convolution formula to
consider the initial state radiation effects[9,10,11,12]:
\begin{eqnarray*}
    \frac{d\sigma}{d\Omega} = \int_{0}^{\Delta}dk\frac{d\sigma^{w}
(s')}{d\Omega}R^{e}_{T}(k), \hskip 5cm (17)
\end{eqnarray*}
where s'=(1-k)s, $\Delta=\frac{E_{\gamma}^{max}}{E_{beam}}$, $\sigma^{w}
(s')$ is the cross section involving the one-loop weak corrections.
The function $R^{e}_{T}$ contains two parts. One is to take
the soft (real and virtual)
photon radiation into account, $S^{e}(\epsilon)$, and the
other is the hard and colinear
photon radiation $H^{e}(k)$[9,10,11,12]:
\begin{eqnarray*}
   R^{e}_{T}(k) = \delta(k)[1+S^{e}(\epsilon)]+\theta(k-\epsilon)H^{e}(k).
\hskip 3cm (18)
\end{eqnarray*}
To the first order, we have
\begin{eqnarray*}
  S^{e}(\epsilon) = \beta_{e}(\ln\epsilon+\frac{3}{4})+\frac{\alpha}
{\pi}e_{e}^{2}(\frac{1}{3}\pi^{2}-\frac{1}{2}) \hskip 4cm (19a)
\end{eqnarray*}
and
\begin{eqnarray*}
  H^{e}(k) = \beta_{e}\frac{1+(1-k)^{2}}{2k}, \hskip 6cm (19b)
\end{eqnarray*}
where
\begin{eqnarray*}
  \beta_{e} = \frac{2\alpha}{\pi} e_{e}^{2}(\ln\frac{s}{m_{e}^{2}}-1)
\hskip 6cm (20a)
\end{eqnarray*}
and
\begin{eqnarray*}
  e_{e} = -1 .\hskip 8.3cm (20b)
\end{eqnarray*}

Numerically we take $M_{Z'}$ as low as possible constrained
by experimental limits, e.g. 395 GeV, 514 GeV and 682 GeV respectively,
and calculate the asymmetry parameter $A^{\mu}_{FB}$ in the two extreme
cases mentioned above, to see the general tendency of the asymmetry varying
with the possible masses of the $Z'$ boson. The values
$\alpha=\frac{1}{137}$, $\sin^{2}{\theta_{w}}$=0.2325, $\epsilon=100 MeV$,
$M_{H^{0}}=200 GeV$ and $M_{t}=150 GeV$ are adopted in the
calculation as well. The results show that the effects to the
asymmetry $A^{\mu}_{FB}$ from pure weak loop corrections are small, roughly
in the order of $10^{-3}$ at the energy region above $Z_{0}$ resonance,
but the QED radiative corrections from the initial state influence the
asymmetry $A^{\mu}_{FB}$ strongly. The curves of forward-backward asymmetry
versus CMS energies $\sqrt S$,
with various $Z'$ masses and $\alpha_{LR}=1.53$
without QED and weak radiative corrections
but in SM and LRM both are plotted in Fig.2. The values of $A^{\mu}_{FB}$
with various $Z'$ masses but $\alpha_{LR}=\sqrt{2/3}$ in LRM at tree level
are nearly the same with that in SM. The discrepancies between those values
are less than $10^{-3}$ in the LEP 200 energy range, and the dependence on
the extra neutral boson $Z'$ mass is not strong.
We may see the fact from Fig.2 that the $Z'$ effects to
$A_{FB}^{\mu}$ is increasing with increasing of $\alpha_{LR}$ and
decreasing of the $Z'$ mass. The $A^{\mu}_{FB}$ versus
$\sqrt{S}$ curves with
QED radiative corrections and weak one-loop corrections for SM and LRM both,
are plotted in Fig.3, when taking $\alpha_{LR}=1.53$, but in Fig.4
when taking
$\alpha_{LR}=\sqrt{2/3}$, respectively.
They keep a similar tendency as
shown in Fig.2. In order to see the
$A_{FB}^{\mu}$ shifts caused by the LRM physics clearly, we define
\begin{eqnarray*}
  \delta A_{FB}^{\mu} = A_{FB}^{\mu,SM}-A_{FB}^{\mu,LRM}  \hskip 6cm (21)
\end{eqnarray*}
and plot it versus the CMS energy
of collision with $M_{Z'}=514 GeV$ in Fig.5.

In summary, we have calculated the forward-backward-asymmetry
of the process
$e^{+} e^{-} \rightarrow \mu^{+} \mu^{-}$ in two extreme cases of LRM
at the tree level and electroweak loop correction levels. Comparisons
on the asymmetry of LRM with those of SM are made precisely.
We find that the QED corrections influence the results strongly comparing
with the weak corrections.
However having them concerned we still may conclude that probably same
indirect indication about the $Z'$ extra boson
predicted by LRM may be obtained through precisely measuring $A^{\mu}_{FB}$,
as long as the measurements on the asymmetry may reach at an accuracy to a
few percents and the mass of the boson $Z'$ is not too heavy. In this paper,
although we focus on the
process $e^{+} e^{-} \rightarrow \mu^{+} \mu^{-}$ only,
for the other processes, such as the process $e^{+} e^{-} \rightarrow
b \bar{b}$, the asymmetry have
a similar tendency in fact[13]. Finally
one thing should be noted is that in the LRM
with $\alpha_{LR}=\sqrt{2/3}$, that
of $E_{6}$ inspired one, the $Z'$ effects to the asymmetry $A^{\mu}_{FB}$
are smaller than that with $\alpha_{LR}=1.53$,
the exact symmetric one of left and right
hand. That is because in
the $\alpha_{LR}=1.53$ case the coupling constant ratio $g_{R}/g_{L}$ is
larger than that in the $\alpha_{LR}=\sqrt{2/3}$ case, so the $Z'$ effect
in $A_{FB}^{\mu}$ is enhanced comparing with that in the later case.

\vskip 2.4cm
{\bf Acknowledgement}\\
This research was supported in part by the Grants of the National
Natural Science Foundation of China under
a special support one and a specific one with
No.19275040, as well as the Grant LWTZ-1298 of Chinese Academy of Science.

\newpage
\noindent
{\Large{\bf Figure Captions}}
\vskip 5mm
\noindent
{\bf Figure 1} All the Feynman diagrams which
contribute to process $e^{+} e^{-}
\rightarrow \mu^{+} \mu^{-}$ at the tree level.\\
\vskip 2mm
\noindent
{\bf Figure 2} The forward-backward-asymmetry $A_{FB}^{\mu}$ versus CMS energy
$\sqrt{s}$ without radiative corrections in SM and LRM (in $\alpha_{LR}=1.53$
case). The full line is of SM; the short-dashed line is of LRM with $M_{Z'}
=395 GeV$;
the dotted line is of LRM with $M_{Z'}=514 GeV$; the long-dashed line is of
LRM with $M_{Z'}=682 GeV$. \\
\vskip 2mm
\noindent
{\bf Figure 3} The forward-backward-asymmetry $A_{FB}^{\mu}$ versus CMS energy
$\sqrt{s}$ with QED radiative correction and one-loop weak corrections in SM
and LRM (in $\alpha_{LR}=1.53$ case).
The full line is of SM; the short-dashed line is of LRM with $M_{Z'}=395 GeV$;
the dotted line is of LRM with $M_{Z'}=514 GeV$; the long-dashed line is of
LRM with $M_{Z'}=682 GeV$. \\
\vskip 2mm
\noindent
{\bf Figure 4} The forward-backward-asymmetry $A_{FB}^{\mu}$ versus CMS energy
$\sqrt{s}$ with QED radiative correction and one-loop weak corrections in SM
and LRM (in $\alpha_{LR}=\sqrt{2/3}$ case).
The full line is of SM; the short-dashed line is of LRM with $M_{Z'}=395 GeV$;
the dotted line is of LRM with $M_{Z'}=514 GeV$; the long-dashed line is of
LRM with $M_{Z'}=682 GeV$. \\
\vskip 2mm
\noindent
{\bf Figure 5} Plot of $\delta A_{FB}^{\mu}=A_{AB}^{SM} - A_{AB}^{LRM}$
 versus $\sqrt{s}$ with $M_{Z'}=514 GeV$
(here the QED radiative correction and one-loop weak corrections are included
in the forward-backward-asymmetry $A_{AB}^{SM}$ and $A_{AB}^{LRM}$).
The full line is for $\alpha_{LR}
=1.53$ and dashed line is for $\alpha_{LR}=\sqrt{2/3}$.

\end{document}